\documentclass{PoS}

\usepackage{threeparttable}
\usepackage{multirow}
\usepackage{amsmath}
\usepackage{amsfonts}
\usepackage{lineno}
\usepackage[utf8]{inputenc}
\title{A search for IceCube events in the direction of ANITA neutrino candidates}

\ShortTitle{IceCube-ANITA Point Source Search}
\author{
The IceCube Collaboration\footnote{For collaboration list, see PoS(ICRC2019) 1177.}\\
{\itshape \href{http://icecube.wisc.edu/collaboration/authors/icrc19_icecube}{http://icecube.wisc.edu/collaboration/authors/icrc19\_icecube}}\\
E-mail: \email{pizzuto@wisc.edu}
}

\abstract{

The Antarctic Impulsive Transient Antenna (ANITA) collaboration has reported a total of three neutrino candidates from the experiment's first three flights. One of these was the lone candidate in a search for Askaryan radio emission, and the others can be interpreted as tau-neutrinos, with important caveats. Among a variety of explanations for these events, they may be produced by astrophysical transients with various characteristic timescales. We test the hypothesis that these events are astrophysical in origin by searching for IceCube counterparts. Using seven years of IceCube data from 2011 through 2018, we search for neutrino point sources using integrated, triggered, and untriggered approaches, and account for the substantial uncertainty in the directional reconstruction of the ANITA events. Due to its large livetime and effective area over many orders of magnitude in energy, IceCube is well suited to test the astrophysical origin of the ANITA events.\\

\vspace{4mm}
{\bfseries Corresponding authors:}
Anastasia Barbano$^{1}$, Teresa Montaruli$^{1}$, \speaker{Alex Pizzuto}$^{2}$, Justin Vandenbroucke$^{2}$\\
{$^{1}$ \itshape D\'epartement de physique nucl\'eaire et corpusculaire, Universit\'e de Gen\`eve, CH-1211 Gen\`eve, Switzerland }\\
{$^{2}$ \itshape Dept. of Physics and Wisconsin IceCube Particle Astrophysics Center, University of Wisconsin, Madison, WI 53706, USA }
}

\FullConference{36th International Cosmic Ray Conference -ICRC2019-\\
		July 24th - August 1st, 2019\\
		Madison, WI, U.S.A.}

\begin{document}

\section{Introduction}
\label{sec:intro}
Despite the detection of high-energy neutrinos of cosmic origin by IceCube in 2013 \cite{Aartsen:2013jdh, Aartsen:2014gkd}, the origin of astrophysical neutrinos remains unknown. Although there is evidence for the first identified neutrino source \cite{IceCube:2018cha}, the overwhelming majority of the measured neutrino flux remains unexplained. 

The Antarctic Impulsive Transient Antenna (ANITA) experiment is a balloon experiment primarily designed to detect the ultra-high energy (UHE) cosmogenic neutrino flux \cite{Hoover:2010qt, Allison:2018cxu} from interactions between UHECRs with the cosmic microwave background (CMB) \cite{Greisen:1966jv, Zatsepin:1966jv}. Among searches for Askaryan emission, one Askaryan candidate (AC) event has been simultaneously identified in one analysis and found to be subthreshold in another, but is noted to have signal shape consistent with impulsive broadband emission that is characteristic of neutrino origin \cite{Allison:2018cxu}. In addition to this event, ANITA has now reported on two events that are consistent with an upgoing astrophysical $\nu_{\tau}$ \cite{Gorham:2016zah, Gorham:2018ydl}. For a complete list of details of these events, see Table \ref{tab:candidates}. In this scenario, a $\nu_{\tau}$ undergoes a charged-current interaction (CC) near the ice-air interface, producing a $\tau$-lepton whose subsequent decay in the atmosphere produces an extensive air shower (EAS). This signal is distinguishable from downward moving cosmic-ray induced EASs, as the radio signals from cosmic-ray induced EASs acquire a phase reversal from reflection off of the Antarctic ice, while an upgoing $\tau$ induced EAS does not display this phase reversal.

\begin{table*}[bht] 
\centering
\begin{threeparttable}
	\caption{Properties of the neutrino candidate events from the first three flights of ANITA, from \cite{Allison:2018cxu, Gorham:2016zah, Gorham:2018ydl}. The two Anomalous ANITA Events (AAE) are those consistent with an upgoing $\nu_{\tau}$ interpretation. Localization uncertainties are expressed as major and minor axis standard deviations, position angle} \label{tab:candidates}
	\begin{tabular}{ l |c |c |c}
	    \hline
	    \hline
	    & \textbf{AAE-061228} & \textbf{AAE-141220} & \textbf{AC-150108} \\ \hline \hline
		Detection Channel & Geomagnetic& Geomagnetic & Askaryan \\
		Date (UTC) & 2006-12-28& 2014-12-20& 2015-01-08 \\
		Time (UTC) & 00:33:20.0 & 08:33:22.5 & 19:04:24.2 \\
		RA, Dec (J2000) & 282$^\circ$.14, +20$^\circ$.33 & 50$^\circ$.78, +38$^\circ$.65 & 171$^\circ$.45, +16$^\circ$.30\\
		Localization Uncertainty & 1$^\circ$.5 $\times$ 1$^\circ$.5, 0$^\circ$.0 & 1$^\circ$.5 $\times$ 1$^\circ$.5, 0$^\circ$.0 & 5$^\circ$.0 $\times$ 1$^\circ$.0, +73$^\circ$.7 \\
		Reconstructed Energy (EeV) & 0.6 $\pm$ 0.4 & $0.56^{+0.30}_{-0.20}$ & $\geq$ 10\\
		Earth Chord Length (km) & 5740 $\pm$ 60 & 7210 $\pm$ 55 & -
	\end{tabular}
	\end{threeparttable}
	\vspace{0.1in}
\end{table*}

However, under Standard Model assumptions, the upgoing neutrino interpretation poses many challenges. First, EeV neutrinos traversing the required chord lengths through the Earth have extremely small survival probabilities \cite{Gorham:2016zah}. Second, if these events are of diffuse flux origin, they would imply fluxes that exceed bounds set by multiple experiments, including self-inconsistencies in ANITA data alone because of the unfavorable arrival directions of the events \cite{Romero-Wolf:2018zxt}. 

If the events from ANITA are considered to have arisen from individual cosmic accelerators, then there is no inconsistency with diffuse flux limits, especially those with short characteristic timescales of emission. Thus, the detection of a single event with EeV energies by ANITA, if created by an $E^{-\gamma}$ power law flux, suggests a large excess of TeV-PeV neutrinos, to which neutrino telescopes such as IceCube would be sensitive. This excess should be especially apparent if the sources which can produce EeV neutrinos have a small source density, while an underestimation of the source density could lead to an overestimation of the expected flux \cite{Strotjohann:2018ufz}. Here, we use IceCube to investigate the hypothesis that the ANITA events were from neutrino sources. 

\section{Data Sample}
\label{sec:icecube}
IceCube is a cubic-kilometer neutrino detector installed in the ice at the geographic South Pole \cite{Aartsen:2016nxy} between depths of 1450 m and 2450 m, completed in 2010. Reconstruction of the direction, energy and flavor of the neutrinos relies on the optical detection of Cherenkov radiation emitted by charged particles produced in the interactions of neutrinos in the surrounding ice or the nearby bedrock.

At the reconstructed directions of the ANITA events, the Earth attenuates the majority of the atmospheric muon flux and the background is dominated by atmospheric neutrinos from cosmic-ray air showers \cite{Haack:2017dxi}. All of these analyses use an event selection which was optimized for point-source searches, and more details of this event sample's properties, such as effective area, are described in \cite{Carver:2019icrc_ps}. For these analyses, we consider data from the full detector configuration of 86 strings and about $8.97\cdot 10^5$ events from 2532 days are analyzed. 

\section{Unbinned likelihood analyses}
\label{sec:methods}
To search for counterparts to ANITA events, we perform three separate analyses that incorporate the information from the ANITA events through a joint likelihood, as described in \cite{Schumacher:2019qdx}. At discrete locations on the sky, $\mathbf{x}_s$, we maximize the likelihood, $\mathcal{L}$, with respect to the parameters $n_s$ and $\mathbf{\alpha}$, given by

\begin{eqnarray} \label{eq:general_likelihood}
    \mathcal{L} = \lambda \prod_{i=1}^{N} \Bigg( \frac{n_s}{n_s + n_b}S(\mathbf{x}_i, \mathbf{x}_s, \mathbf{\alpha})  + \frac{n_b}{n_s + n_b}B(\mathbf{x}_i, \mathbf{x}_s) \Bigg) P_A(\mathbf{x}_s) ,
\end{eqnarray}
where $n_b$ is the expected number of background events, $n_s$ is the fitted number of signal events, $\{\mathbf{x}_i\}$ are the reconstructed event properties of IceCube neutrino candidates, and $P_A$ is the spatial probability distribution function (PDF) characterizing the uncertainty on the reconstruction of the ANITA events. $\alpha$ represents any additional free parameters which are fit for, such as the best fit spectral index, $\gamma$, and are later described for each individual analysis. $B$ is a background PDF of energy, direction, and time, and is parameterized from data. The signal hypotheses are encapsulated by $\lambda$ and $S$, which describe the number of events on the sky and the reconstructed properties of such events, respectively, and are unique to each individual analysis. This technique of including ANITA information through a prior is employed in three different search strategies: \textit{prompt}, \textit{rolling}, and \textit{steady}. 

\begin{figure}[t!]
\centering
\vspace{-0.6cm}
    \includegraphics[width=0.99\textwidth,trim={0 3cm 0 0 }]{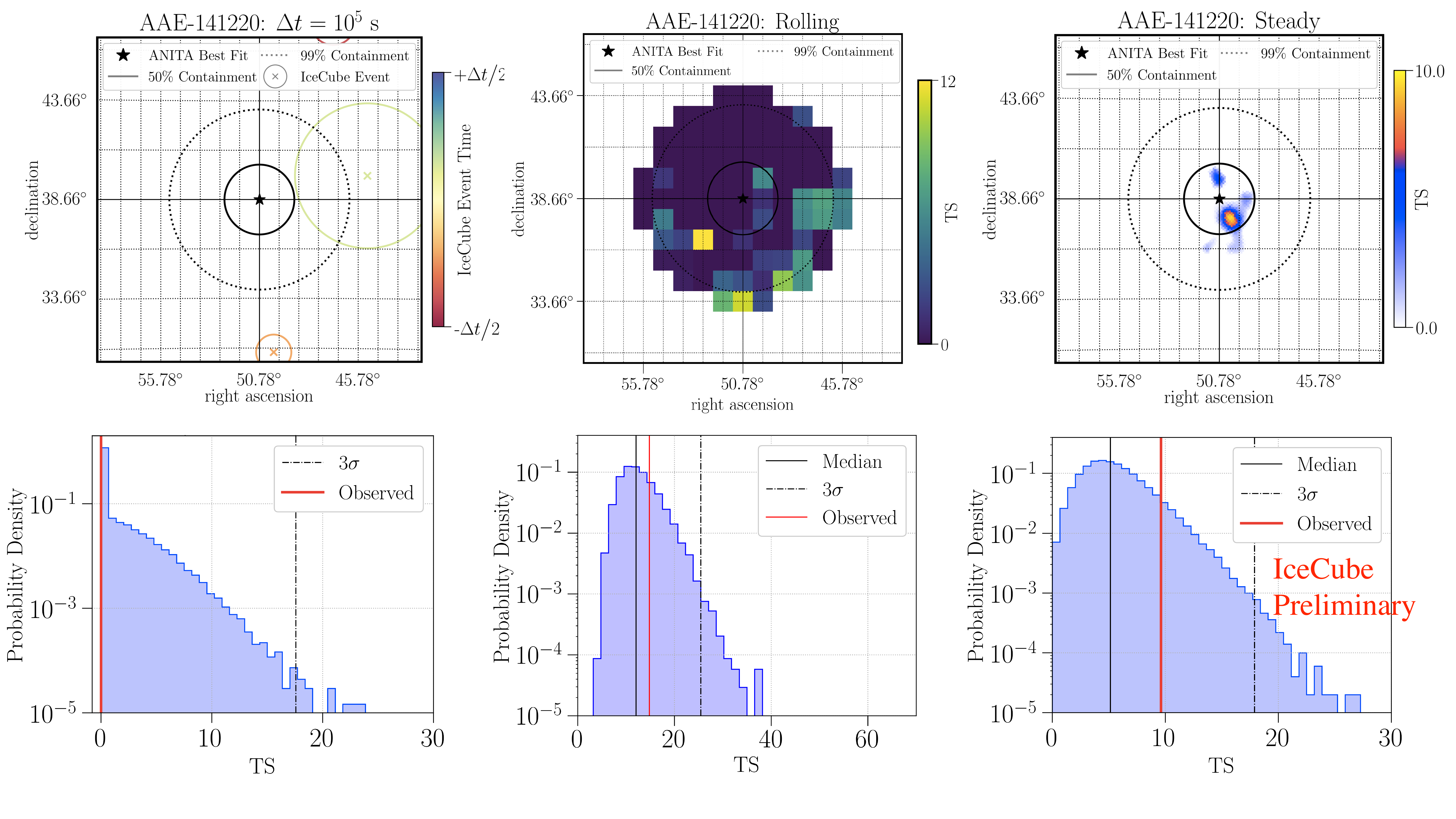}
\caption{Skymaps (top) and TS distributions (bottom) for AAE-141220 for the prompt (left), steady (middle), and rolling (right) analyses. Observed TS values (shown in red) are compared to distributions from time-scrambled data realizations to quantify the significance.} \label{fig:skymaps}
\end{figure}

\subsection{Prompt}
\label{sub:transient}
The first analysis searches for spatial and temporal coincidence of IceCube and ANITA neutrino events by only considering time windows, $\Delta t$, centered on each ANITA event, which we call the \textit{on-time window}. To help distinguish potential signals for time windows in which the expected number of background events is small, we set 
\begin{equation}
    \lambda = \frac{(n_s + n_b)^N}{N!}\cdot {\rm e} ^{-(n_s + n_b)}
\end{equation} 
as in \cite{Aartsen:2017zvw, Aartsen:2014aqy}. The likelihood is only maximized with respect to $n_s$ and the energy dependence in $S$ is fixed to an $E^{-2}$ spectrum, culminating in the definition of a test statistic (TS), which arises from the logarithmic likelihood ratio between the best-fit likelihood and that of the null hypothesis,
\begin{equation} \label{eq:transient_TS}
    \text{TS} = - 2\;n_s + \sum_{i=1}^{N}2\;\log \left[ 1 + \frac{n_s S(\mathbf{x}_i, \mathbf{x}_s)}{n_b B(\mathbf{x}_i)} \right] + 2 \log \left[ \frac{P_A (\mathbf{x}_s)}{P_A (\mathbf{x}_0)} \right] \; ,
\end{equation}
where $\mathbf{x}_0$ is the reported best fit location of the ANITA event. TS is calculated for all $\mathbf{x}_s$, and the maximum value is reported. We perform a model-independent search by separately considering constant emission over various time windows for each of the ANITA events, similar to previous IceCube searches for gamma-ray bursts and fast radio bursts \cite{Aartsen:2017zvw, Aartsen:2014aqy}. Here, we consider time windows of 10 s, $10^3$ s, $10^5$ s. AAE-061228 is excluded from this analysis because it occurred before the start of our event selection. For AAE-141220, we only consider the largest two windows, as IceCube was temporarily not collecting data at the time of the event, due to a run transition that had begun approximately 0.5 seconds before the event and lasting for about one minute.

\subsection{Rolling}
\label{sub:flare}
The second analysis searches for spatial and temporal clustering of IceCube events in agreement with the ANITA event PDFs, but does not require the IceCube events to be temporally coincident with the ANITA events \cite{Braun:2009wp,Aartsen:2015wto}. Here, the signal PDF is defined as:
\begin{eqnarray} \label{eq:SigPDFUntrig}
    S = P_i^{signal}(\mathbf{x}_i, \mathbf{x}_s, \mathbf{\sigma}) \cdot \epsilon_i^{signal}(E_i,\delta_i,\gamma) \cdot T_i^{signal},
\end{eqnarray}
where the three terms are a spatial, energy and a time PDF, respectively.
The spatial term evaluates the probability of an event originating from a certain location according to a two-dimensional Gaussian function with best-fit direction and angular resolution $\mathbf{x}_i$ and $\sigma_i$, respectively. The energy PDF $\epsilon_i^{signal}(E_i,\delta_i,\gamma)$ 
describes the probability of obtaining a reconstructed energy $E_i$
for an event produced by a source at declination $\delta_i$ of a given 
$E^{-\gamma}$ power-law energy spectrum. The time term assumes signal events are distributed according to a one-dimensional Gaussian function in time, with mean and standard deviation $t_0$ and $\sigma_t$. 
The TS is defined as:
\begin{eqnarray} \label{eq:TSUntrig}
    \text{TS}(\hat{n_s},\hat{\gamma},\hat{\sigma_t},\hat{t_0}; \hat{\mathbf{x}_s}) = -2 \, \text{log} \Bigg[\frac{\text{T}}{\sqrt{2\pi}\hat{\sigma_t}} \times \frac{\mathcal { L } (n_s=0)}{\mathcal{L}(\hat{n_s},\hat{\gamma},\hat{\sigma_t},\hat{t_0})}\Bigg],
\end{eqnarray}
where $\hat{n_s},\hat{\gamma},\hat{\sigma_t},\hat{t_0}, \hat{\mathbf{x}_s}$ are the best-fit values
from the TS maximization and $T$ is the total livetime of the data-taking period. 
The term that multiplies the likelihood function ratio in Eq.~\ref{eq:TSUntrig} 
is a marginalization term to avoid undesired biases toward finding short flares.
The TS($\hat{n_s},\hat{\gamma},\hat{\sigma_t},\hat{t_0}$) is calculated at every degree in declination and right ascension, beginning from the central 
coordinates of the ANITA events and covering 3.5$\sigma$ of the their 
two-dimensional spatial PDFs. 
The position in the grid corresponding to the maximum value of TS is used
as a seed to perform a further TS($\hat{n_s},\hat{\gamma},\hat{\sigma_t},\hat{t_0}, \hat{\mathbf{x}_s}$) maximization, where also the direction $\mathbf{x}_s$ of the 
source is reconstructed. An example of the TS maximization over the 
defined sky grid is shown in the top-right panel of Fig.~\ref{fig:skymaps} for AAE-141220. The red dashed lines indicate the position of the most significant point-like source
as reconstructed from the likelihood fit.

\subsection{Steady}
\label{sub:steady}
A final analysis is performed to test for spatial clustering over the considered seven years of data. As the total number of events is large, we set $\lambda$ to 1, and the increase in statistics allows us to fit for $\gamma$ in the range $1 \leq \gamma \leq 4$ in addition to $n_s$, as is done in many previous IceCube analyses \cite{Aartsen:2016tpb, Aartsen:2016oji, Aartsen:2013uuv, Abbasi:2010rd, Aartsen:2014cva}, and using the same energy PDF, $\epsilon_i^{signal}(E_i,\delta_i,\gamma)$, that is used in the rolling analysis. In the regime where the expected number of background events is large, the ratio of $(n_s + n_b) / N$ approaches unity, so we explicitly set $n_s + n_b$ to be equal to the number of events, $N$, and at all $\mathbf{x}_s$ we calculate the redefined TS
\begin{equation}
    \text{TS} = 2 \cdot \log \left[ \frac { \mathcal { L } \left( \mathbf { x } _ { s } , \hat { n } _ { s } , \hat { \gamma } \right) } { \mathcal { L } \left( \mathbf { x } _ { s } , n _ { s } = 0 \right) } \right] + 2 \log \left[ \frac{P_A (\mathbf{x}_s)}{P_A (\mathbf{x}_0)} \right]\; , \qquad 
\end{equation}
with best-fit values $\hat{n}_s$ and $\hat{\gamma}$. 

\section{Results}
\label{sec:results}

No significant correlation is found in any of the analyses above the expectation from background. The p-values cited in this section are calculated from pseudo-experiments generated from experimental data. In these realizations, events are assigned a random time while maintaining their directional reconstruction in local detector coordinates, which effectively scrambles the events in both right ascension and time \cite{Aartsen:2015wto}. The most significant observation results from the time-integrated search for AAE-141220, with a p-value of 0.08, which we find to be completely consistent with background. 

Figure \ref{fig:skymaps} displays the unblinded skymaps for the prompt, steady, and rolling analyses from left to right in the top panels for AAE-141220. Bottom panels of Figure \ref{fig:skymaps} show the comparison of the observed TS values for each analysis, at the position of the red lines, to their respective TS distributions from pseudo-experiments using time-scrambled data.

\begin{figure}
\centering
  \makebox[\textwidth]{\makebox[1.00\textwidth]{
  \begin{minipage}{0.48\textwidth}
  \centering
  \includegraphics[width=\textwidth, trim = {2cm 0.5cm 0cm 2cm}]{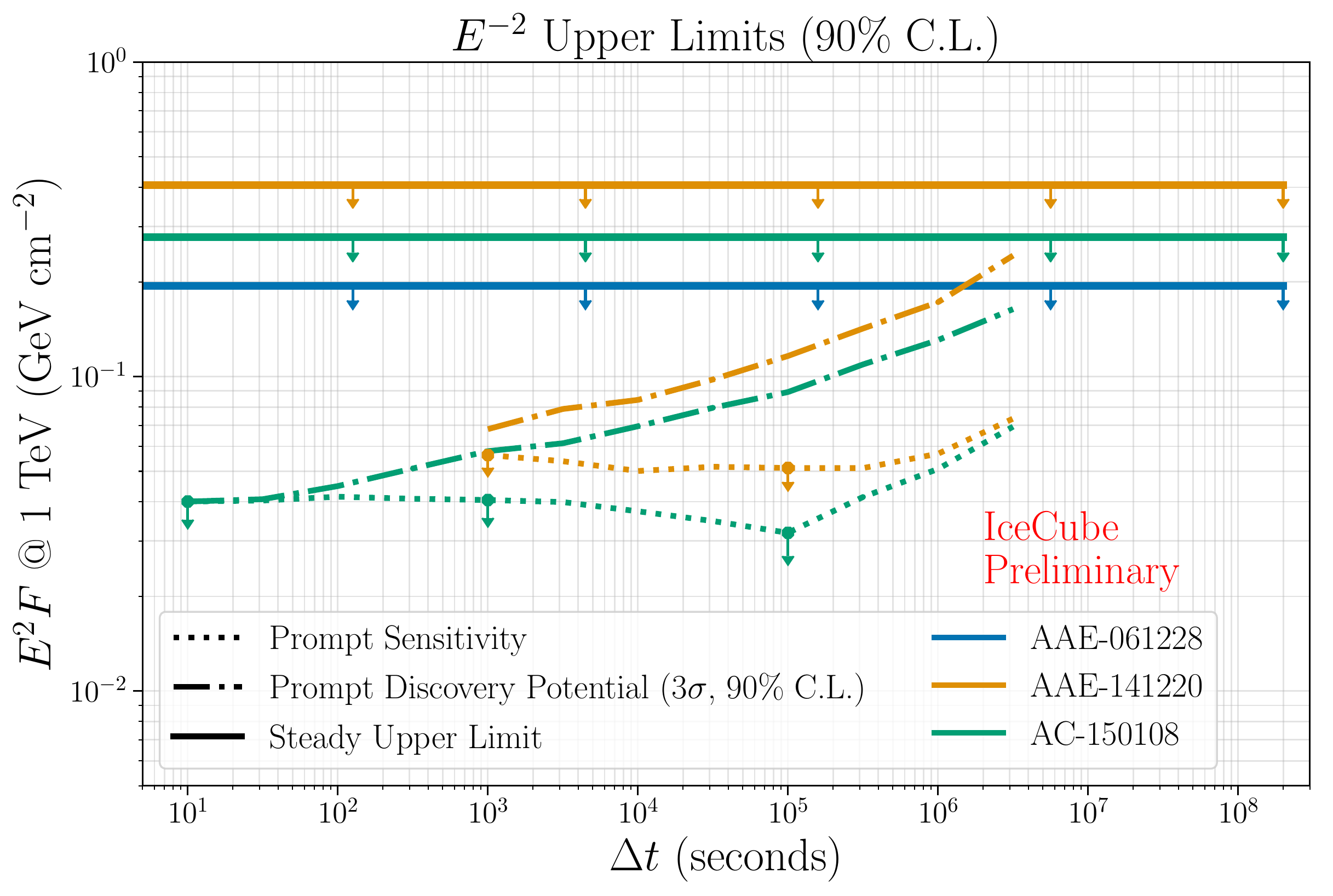}
  \caption{Sensitivity and upper limits (90\% confidence level) on the flux normalization for an $E^{-2}$ source spectrum as a function of $\Delta t$ from the prompt analysis, compared to the upper limits (solid) from the time-integrated analysis. The central 90\% intervals of the expected neutrino energies for these spectra are 1TeV-1PeV. }
  \label{fig:upper_limit}
  \end{minipage}
  \hfill
  \begin{minipage}{0.48\textwidth}  
  \centering 
  \includegraphics[width=\textwidth, trim = {1cm 0.5cm 1cm 1cm}]{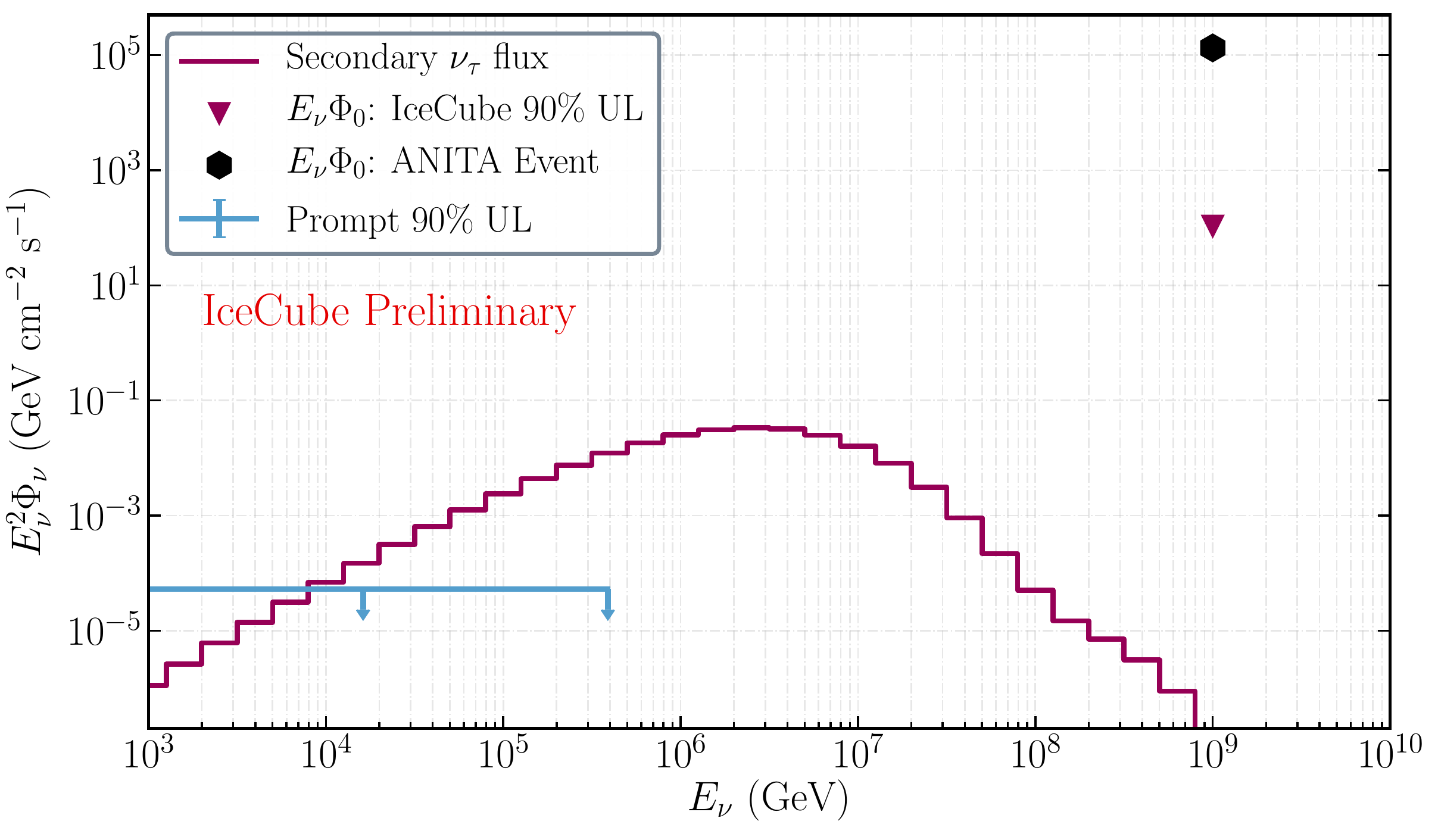}
  \caption{Upper limits (90\% C.L.) placed by calculating the secondary neutrino flux (purple histogram) from an incident flux of EeV neutrinos assuming constant emission over $10^3$ s and comparing to the non-observation of IceCube events in the prompt analysis. The flux implied by the ANITA observations (black) using information about ANITA's acceptance \cite{Romero-Wolf:2018zxt} overshoots this upper limit (purple arrow) by many orders of magnitude. For comparison, upper limits on the muon-neutrino flux from the prompt analysis are shown in blue.}
  \label{fig:taurunner_limits}
  \end{minipage}}}
  \vspace{0.2in}
\end{figure}

In the absence of a significant signal, upper limits (90\% confidence level) for the time-integrated flux are set for each ANITA event using the triggered and time-integrated analyses (Figure \ref{fig:upper_limit}). Limits provided in Table \ref{tab:results} are set assuming an $E^{-2}$ neutrino spectrum and that the source is in any location consistent with the ANITA PDF. 

%\begin{figure}[t!]
%\centering
%    \includegraphics[width=0.65\textwidth]{figures/ANITA_transient_Upper_limits.png}
%\caption{Sensitivity and upper limits (90\% confidence level) on the flux normalization for an $E^{-2}$ source spectrum as a function of $\Delta t$ from the prompt analysis, compared to the upper limits (solid) from the time-integrated analysis. The central 90\% intervals of the expected neutrino energies for these spectra are 1TeV-1PeV. } \label{fig:upper_limit}
%\end{figure}

\begin{table*}[htb!]
    \caption{Analysis results and upper limits. 
    %The time windows for the untriggered analysis are written as the offset from ANITA detection time, best-fit flare duration ($t_0 - T$, $\sigma_t$). 
    Upper limits (90\% C.L) are on the time-integrated $E^{-2}$ power law flux from a point-source following the spatial probability distribution provided by ANITA. \vspace{0.2cm}}
    \centering
    \begin{tabular}{l | c| c| c | c} \hline
     Event & Analysis & Time Window  & $p$-value & Upper limit (GeV $\cdot$ cm$^{-2}$)  \\ \hline \hline
     \multirow{3}{*}{AAE-061228} & Steady & IC86-I - IC86-VII & 0.606 &  0.195 \\ \cline{2-5}
     & \multirow{2}{*}{Rolling} & IC86-I & 0.562 & - \\ 
     & & IC86-II - IC86-VII & 0.208 & - \\ \hline
     \multirow{5}{*}{AAE-141220} & & 10s & 1.0 & - \\ 
	 & Prompt & $10^3$s & 1.0 & 0.053 \\ 
	 & & $10^5$s & 1.0 & 0.051 \\ \cline{2-5}
	 & Steady & IC86-I - IC86-VII & 0.081 & 0.401 \\ \cline{2-5}
	 & \multirow{2}{*}{Rolling} & IC86-I & 0.342 & - \\ 
     & & IC86-II - IC86-VII & 0.224 & - \\ \hline
	 \multirow{5}{*}{AC-150108} & & 10s & 1.0 & 0.040 \\ 
	 & Prompt & $10^3$s & 1.0 & 0.041 \\ 
	 & & $10^5$s & 1.0 & 0.032 \\ \cline{2-5}
	 & Steady & IC86-I - IC86-VII & 0.210 & 0.278 \\ \cline{2-5}
	 & \multirow{2}{*}{Rolling} & IC86-I & 0.636 & - \\ 
     & & IC86-II - IC86-VII & 0.512 & - \\ \hline
    \end{tabular}
    \label{tab:results}
\end{table*}

\vspace{-0.4in}
\section{Discussion}
\label{sec:discussion}
For many astrophysical sources, power law spectra in photons are common, lending credibility to the choice of testing power laws for corresponding neutrino spectra. For this analysis, interpolating between the energy range at which IceCube is sensitive and that of ANITA may not be justified.

However, the limits we set for TeV-PeV energies are still constraining for fluxes at EeV energies. As is shown in \cite{Safa:2019icrc_ps}, any incident flux with an EeV $\nu_{\tau}$ component that traverses the Earth will result in a secondary flux of lower energy neutrinos of all flavors, to which IceCube would be sensitive. Here, we analyze how constraining our limits are on any point source flux that includes EeV neutrinos using the \texttt{TauRunner} code and prescription described in \cite{Safa:2019icrc_ps}. We inject a flux of EeV tau neutrinos ($\Phi = \phi_0 \delta (E - E_0)$ with $E_0=$ 1 EeV) at an angle corresponding to that of AAE-141220, and find the spectral shape of the secondary neutrino flux. Using the non-observation of coincident events from the prompt analysis for a time window of $10^3$ s, we find the maximum allowed flux normalization (with a 90\% C.L.) for this incident flux, by comparing the cascaded secondary flux to an observation of zero events by IceCube. The results are displayed in Figure \ref{fig:taurunner_limits}. Although IceCube's sensitivity is peaked many orders of magnitude below the reconstructed energies of the ANITA events, the limits set on any potential neutrino source that created AAE-141220 are constraining.

\section{Conclusion}
\label{sec:conclusion}
Some of the recent detections of neutrino events by ANITA are considered anomalous due to the small survival probability of EeV neutrinos traversing long chord lengths. For the non-anomalous AC event, we have placed upper limits on the neutrino emission from a point source whose location is distributed according to the event's PDF. Additionally, for the AAE, the limits placed on point source emission are below the implied fluxes, as long as the source density which could have produced these events is small. 

These new limits, in conjunction with the inconsistency of diffuse interpretations, could require a non-astrophysical interpretation of the AAE. While numerous explanations incite physics beyond the Standard Model \cite{Cherry:2018rxj, Connolly:2018ewv, Fox:2018syq, Esteban:2019hcm}, it has recently been suggested that the AAE could be explained by downward-going CR-induced EAS that reflected off of subsurface features in the Antarctic ice \cite{Shoemaker:2019xlt} or from coherent transition radiation from cosmic-ray air showers \cite{deVries:2019gzs}. Regardless of the origins of the AAE, these studies highlight the necessity to have a deeper understanding of the Antarctic ice for next generation neutrino and cosmic-ray experiments alike.

\bibliographystyle{ICRC}
\bibliography{references}

\providecommand{\href}[2]{#2}\begingroup\raggedright\begin{thebibliography}{10}

\bibitem{Aartsen:2013jdh}
{\bf IceCube} Collaboration, M.~G. Aartsen et~al., {\em Science} {\bf 342}
  (2013) 1242856.

\bibitem{Aartsen:2014gkd}
{\bf IceCube} Collaboration, M.~G. Aartsen et~al., {\em Phys. Rev. Lett.} {\bf
  113} (2014) 101101.

\bibitem{IceCube:2018cha}
{\bf IceCube} Collaboration, M.~G. Aartsen et~al., {\em Science} {\bf 361}
  (2018) 147--151.

\bibitem{Hoover:2010qt}
{\bf ANITA} Collaboration, S.~Hoover et~al., {\em Phys. Rev. Lett.} {\bf 105}
  (2010) 151101.

\bibitem{Allison:2018cxu}
{\bf ANITA} Collaboration, P.~W. Gorham et~al., {\em Phys. Rev.} {\bf D98}
  (2018) 022001.

\bibitem{Greisen:1966jv}
K.~Greisen, {\em Phys. Rev. Lett.} {\bf 16} (1966) 748--750.

\bibitem{Zatsepin:1966jv}
G.~T. Zatsepin and V.~A. Kuzmin, {\em JETP Lett.} {\bf 4} (1966) 78--80. [Pisma
  Zh. Eksp. Teor. Fiz.4,114(1966)].

\bibitem{Gorham:2016zah}
{\bf ANITA} Collaboration, P.~W. Gorham et~al., {\em Phys. Rev. Lett.} {\bf
  117} (2016) 071101.

\bibitem{Gorham:2018ydl}
{\bf ANITA} Collaboration, P.~W. Gorham et~al., {\em Phys. Rev. Lett.} {\bf
  121} (2018) 161102.

\bibitem{Romero-Wolf:2018zxt}
A.~Romero-Wolf et~al., \href{http://arxiv.org/abs/1811.07261}{{\tt
  arXiv:1811.07261}}.

\bibitem{Strotjohann:2018ufz}
N.~L. Strotjohann, M.~Kowalski, and A.~Franckowiak, {\em Astron. Astrophys.}
  {\bf 622} (2019) L9.

\bibitem{Aartsen:2016nxy}
{\bf IceCube} Collaboration, M.~G. Aartsen et~al., {\em JINST} {\bf 12} (2017)
  P03012.

\bibitem{Haack:2017dxi}
{\bf IceCube} Collaboration, C.~Haack and C.~Wiebusch,  \pos{PoS(ICRC2017)1005}
  (2018).

\bibitem{Carver:2019icrc_ps}
{\bf IceCube} Collaboration,  \pos{PoS(ICRC2019)851} (these proceedings).

\bibitem{Schumacher:2019qdx}
{\bf ANTARES, IceCube, Pierre Auger, Telescope Array} Collaboration,
  L.~Schumacher, {\em EPJ Web Conf.} {\bf 207} (2019) 02010.

\bibitem{Aartsen:2017zvw}
{\bf IceCube} Collaboration, M.~G. Aartsen et~al., {\em Astrophys. J.} {\bf
  857} (2018) 117.

\bibitem{Aartsen:2014aqy}
{\bf IceCube} Collaboration, M.~G. Aartsen et~al., {\em Astrophys. J.} {\bf
  805} (2015) L5.

\bibitem{Braun:2009wp}
J.~Braun, et~al., {\em Astropart. Phys.} {\bf 33} (2010) 175--181.

\bibitem{Aartsen:2015wto}
{\bf IceCube} Collaboration, M.~G. Aartsen et~al., {\em Astrophys. J.} {\bf
  807} (2015) 46.

\bibitem{Aartsen:2016tpb}
{\bf IceCube} Collaboration, M.~G. Aartsen et~al., {\em Astrophys. J.} {\bf
  824} (2016) L28.

\bibitem{Aartsen:2016oji}
{\bf IceCube} Collaboration, M.~G. Aartsen et~al., {\em Astrophys. J.} {\bf
  835} (2017) 151.

\bibitem{Aartsen:2013uuv}
{\bf IceCube} Collaboration, M.~G. Aartsen et~al., {\em Astrophys. J.} {\bf
  779} (2013) 132.

\bibitem{Abbasi:2010rd}
{\bf IceCube} Collaboration, R.~Abbasi et~al., {\em Astrophys. J.} {\bf 732}
  (2011) 18.

\bibitem{Aartsen:2014cva}
{\bf IceCube} Collaboration, M.~G. Aartsen et~al., {\em Astrophys. J.} {\bf
  796} (2014) 109.

\bibitem{Safa:2019icrc_ps}
I.~Safa et~al., {\em PoS} {\bf ICRC2019} (these proceedings).

\bibitem{Cherry:2018rxj}
J.~F. Cherry and I.~M. Shoemaker, {\em Phys. Rev.} {\bf D99} (2019) 063016.

\bibitem{Connolly:2018ewv}
A.~Connolly, P.~Allison, and O.~Banerjee,
  \href{http://arxiv.org/abs/1807.08892}{{\tt arXiv:1807.08892}}.

\bibitem{Fox:2018syq}
D.~B. Fox, et~al., {\em Submitted to: Phys. Rev. D} (2018).

\bibitem{Esteban:2019hcm}
I.~Esteban, et~al., \href{http://arxiv.org/abs/1905.10372}{{\tt
  arXiv:1905.10372}}.

\bibitem{Shoemaker:2019xlt}
I.~M. Shoemaker, et~al., \href{http://arxiv.org/abs/1905.02846}{{\tt
  arXiv:1905.02846}}.

\bibitem{deVries:2019gzs}
K.~D. de~Vries and S.~Prohira, \href{http://arxiv.org/abs/1903.08750}{{\tt
  arXiv:1903.08750}}.

\end{thebibliography}\endgroup

\end{document}